\documentclass{SciPost}

\hypersetup{
    colorlinks,
    linkcolor={red!50!black},
    citecolor={blue!50!black},
    urlcolor={blue!80!black}
}

\usepackage[bitstream-charter]{mathdesign}
\usepackage{subcaption}
\usepackage{fancyhdr}
\usepackage{caption}
\DeclareSymbolFont{usualmathcal}{OMS}{cmsy}{m}{n}
\DeclareSymbolFontAlphabet{\mathcal}{usualmathcal}

\fancypagestyle{SPstyle}{
\fancyhf{}
\lhead{\colorbox{scipostdeepblue}{\bf \color{white} ~SciPost Physics Proceedings }}
\rhead{{\bf \color{scipostdeepblue} ~Submission }}
\renewcommand{\headrulewidth}{1pt}
\fancyfoot[C]{\textbf{\thepage}}
}

\begin{document}

\pagestyle{SPstyle}

\begin{center}{\Large \textbf{\color{scipostdeepblue}{
Autoencoder-based time series anomaly detection for ATLAS Liquid Argon calorimeter data quality monitoring\\
}}}\end{center}

\begin{center}\textbf{
% Use (full) first name (+ middle name initials) + surname format.
% Separate subsequent authors by a comma, omit comma and use "and" for the last author.
% Mark the corresponding author(s) with a superscript symbol in this order
% \star, \dagger, \ddagger, \circ, \S, \P, \parallel, ...
Vilius \v{C}epaitis\textsuperscript{1$\star$},
on behalf of the ATLAS collaboration
}\end{center}

\begin{center}
{\bf 1} University of Geneva, Switzerland
\\[\baselineskip]
$\star$ \href{mailto:vilius.cepaitis@cern.ch}{\small vilius.cepaitis@cern.ch}
\end{center}

\definecolor{palegray}{gray}{0.95}
\begin{center}
\colorbox{palegray}{
  \begin{tabular}{rr}
  \begin{minipage}{0.37\textwidth}
    \includegraphics[width=60mm]{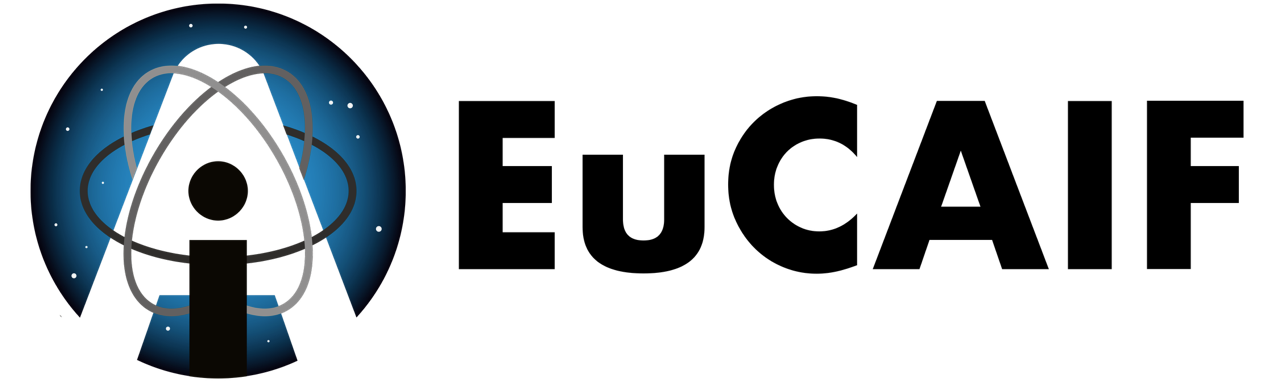}
  \end{minipage}
  &
  \begin{minipage}{0.5\textwidth}
    \vspace{5pt}
    \vspace{0.5\baselineskip} 
    \begin{center} \hspace{5pt}
    {\it The 2nd European AI for Fundamental \\Physics Conference (EuCAIFCon2025)} \\
    {\it Cagliari, Sardinia, 16-20 June 2025
    }
    \vspace{0.5\baselineskip} 
    \vspace{5pt}
    \end{center}
    
  \end{minipage}
\end{tabular}
}
\end{center}

\section*{\color{scipostdeepblue}{Abstract}}
\textbf{\boldmath{
The ATLAS experiment at the LHC employs comprehensive data quality monitoring procedures to ensure high-quality physics data. This contribution presents a long short-term memory autoencoder-based algorithm for detecting anomalies in ATLAS Liquid Argon calorimeter data, represented as multidimensional time series of statistical moments of energy cluster properties. Trained on good-quality data, the model identifies anomalous intervals. Validation is performed using a known short-term issue of noise bursts, and the potential for broader application to transient calorimeter issues is discussed.
}}
\vspace{\baselineskip}

%%%%%%%%%% BLOCK: Copyright information
% This block will be filled during the proof stage, and finilized just before publication.
% It exists here only as a placeholder, and should not be modified by authors.
\noindent\textcolor{white!90!black}{%
\fbox{\parbox{0.975\linewidth}{%
\textcolor{white!40!black}{\begin{tabular}{lr}%
  \begin{minipage}{0.6\textwidth}%
    {\small Copyright attribution to authors. \newline
    This work is a submission to SciPost Phys. Proc. \newline
    License information to appear upon publication. \newline
    Publication information to appear upon publication.}
  \end{minipage} & \begin{minipage}{0.4\textwidth}
    {\small Received Date \newline Accepted Date \newline Published Date}%
  \end{minipage}
\end{tabular}}
}}
}
%%%%%%%%%% BLOCK: Copyright information
% Apply the footer only to this page
\fancypagestyle{firstpage}{%
  \fancyhf{} % clear header and footer
  \fancyfoot[L]{\scriptsize Copyright 2025 CERN for the benefit of the ATLAS Collaboration. CC-BY-4.0 license.}
  \renewcommand{\headrulewidth}{0pt}
  \renewcommand{\footrulewidth}{0.4pt}
}

\thispagestyle{firstpage} % apply only to this page

% For convenience during refereeing we turn on line numbers:
%\linenumbers

% Write your article contents here, starting from first \section.
% An example structure is given below.

\section{Introduction}
\label{sec:intro}

The Liquid Argon (LAr) calorimeter~\cite{ATLAS-TDR-02, ATLAS-TDR-22} is a key component of the ATLAS~\cite{PERF-2007-01} detector at the CERN LHC. It is designed to precisely measure electromagnetic shower energies, such as those originating from electrons and photons. Reliable data taking with many channels over decades necessitates identification of occasional issues. In these proceedings, a generic, proof-of-concept time series anomaly detection (AD) method for the LAr calorimeter data quality monitoring (DQM) is presented. While the LAr calorimeter has excellent data quality efficiency~\cite{DAPR-2018-01}, there are several reasons for complementing the existing infrastructure with such a technique. First, existing algorithms focus on well-known detector issues, whereas an AD algorithm could act as an early warning system for rare, previously unseen problems. Furthermore, such a technique could serve in reducing the substantial human effort required to perform routine DQM checks.

%\begin{figure}[htb!]
%    \centering
%    \includegraphics[width=0.4\linewidth]{LAr_Labeled.png}
%    \caption{Cut-away view of the ATLAS LAr calorimeter, taken from Ref.~\cite{ATLAS:2010blk}.}
%    \label{fig:calorimeter}
%\end{figure}

\section{Methodology}
This section summarises the inputs to the autoencoder-based anomaly detection algorithm as well as the model's architecture and training setup. Additional details of the training procedure can be found in Ref.~\cite{ATL-DAPR-PUB-2024-002}.

\subsection{Liquid argon calorimeter inputs}

\begin{figure}[htb!]
    \centering
    \includegraphics[width=0.8\linewidth]{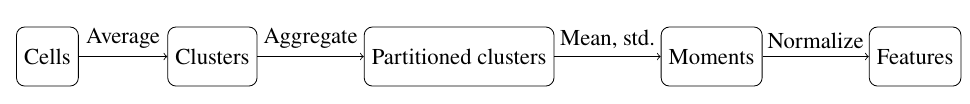}
    \caption{A sketch of the data processing pipeline}
    \label{fig:data-processing-pipeline}
\end{figure}

A sketch of the data processing pipeline is shown in Figure~\ref{fig:data-processing-pipeline}. The whole calorimeter data is represented as a 16-dimensional time series, corresponding to different properties of the ensemble of calorimeter signals in each of the four electromagnetic detector partitions: EMBA, EMBC, EMECA, and EMECC.

The time series data is obtained in a staged approach. First, individual calorimeter signals (cells) are clustered into topo-clusters~\cite{PERF-2014-07}, a standard method used in ATLAS for hadronic object reconstruction. Two topo-cluster variables are used: the \(Q\)-factor and the timing, \(\tau\), calculated based on the pulse shape properties of the corresponding cells. These variables were chosen for their sensitivity to different detector issues. For each topo-cluster variable, two statistical moments (mean, standard deviation) are obtained. Finally, the topo-clusters are themselves aggregated over different physical calorimeter partitions, transforming the variable number of calorimeter signals into a fixed-size representation in the time series. As a final step of data preprocessing, each dimension is normalized to ensure it contributes a similar weight to the reconstruction loss, as discussed in the following section. The time series data represents an irregularly sampled event stream, with entries corresponding to events triggered by ATLAS during roughly one-minute intervals of data-taking.

\subsection{Anomaly detection algorithm}

\begin{figure}[htb]
    \centering
    \includegraphics[width=0.6\linewidth]{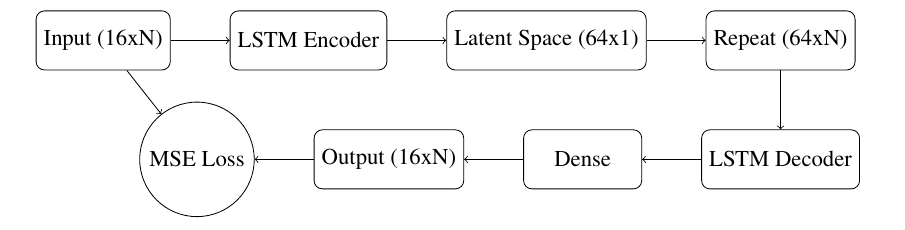}
    \caption{The LSTM autoencoder architecture.}
    \label{fig:autoencoder-lstm}
\end{figure}

An unsupervised autoencoder-based anomaly detection algorithm~\cite{Hinton_Salakhutdinov_2006} is used to identify anomalous time periods. The autoencoder (AE) architecture is shown in Figure~\ref{fig:autoencoder-lstm}. The AE consists of an encoder \(f\) and a decoder \(g\), which compress and reconstruct the input as well as possible. A long short-term memory (LSTM)~\cite{10.1162/neco.1997.9.8.1735} block is used for both the encoder and decoder architecture. The mean square error loss function is minimized between the input time series \(\vec{X}\) and their auto-encoder replicas during training:

\begin{equation}\label{eq:mse-loss}
    \mathcal{L} = \frac{1}{N}\sum_{i=1}^{N}\|\vec{X}_i - g(f(\vec{X}_i))\|^2.
\end{equation}

Several samples of ATLAS data were used for training the autoencoder, corresponding to both proton--proton and heavy--ion data, collected in different years. This combination was performed to obtain a more robust model as the detector, beam, and data collection conditions all changed. Only data certified as good for physics analyses by ATLAS was used in the training.

The algorithm exploits the fact that the detector generally operates well, only occasionally interrupted by brief periods of anomalous activity. The AE aims to reconstruct the bulk of the good data training distribution. LAr detector issues are generally rare, for instance the corresponding data loss during 2015--2018 data-taking was below 1\%~\cite{DAPR-2018-01}. Therefore, such data will be more likely to be poorly reconstructed as an outlier, i.e., to have a higher reconstruction loss. This approach complements conventional DQM techniques within ATLAS, which are based on monitoring distributions in the form of histograms.

\section{Results}

Several independent datasets were employed to evaluate the model performance. To validate the reconstruction of the AE and identify a threshold loss for flagging events as anomalous, a data-taking run heavily enriched in noise bursts was used. 

\begin{figure}[htb!]
    \centering
    \begin{subfigure}[b]{0.45\linewidth}
        \centering
        \includegraphics[height=5cm]{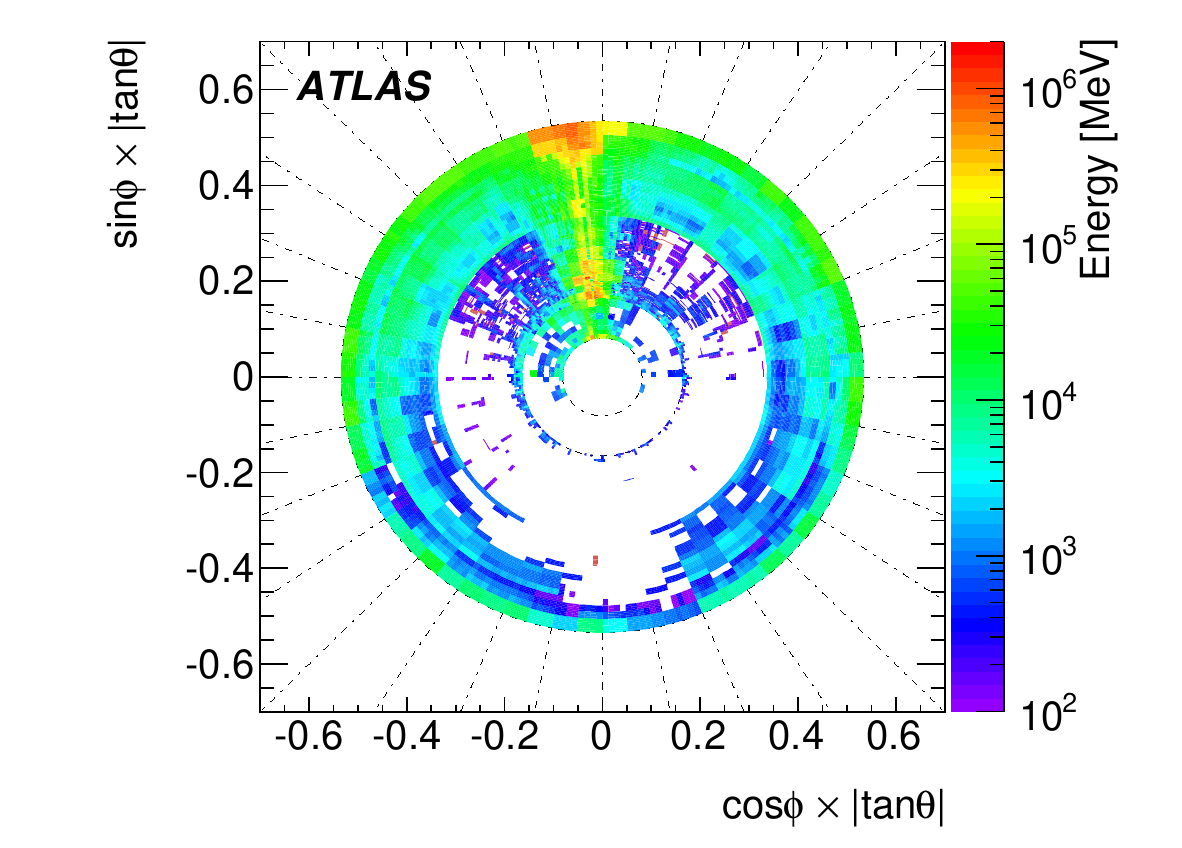}
        \caption{A typical LAr calorimeter noise burst event, corresponding to an energy deposit of a few TeV, shown in the transverse plane. Figure from Ref.~\cite{LARG-2013-01}.}
        \label{fig:noise-burst}
    \end{subfigure}
    \hspace{0.5cm}
    \begin{subfigure}[b]{0.5\linewidth}
        \centering
        \includegraphics[height=5cm]{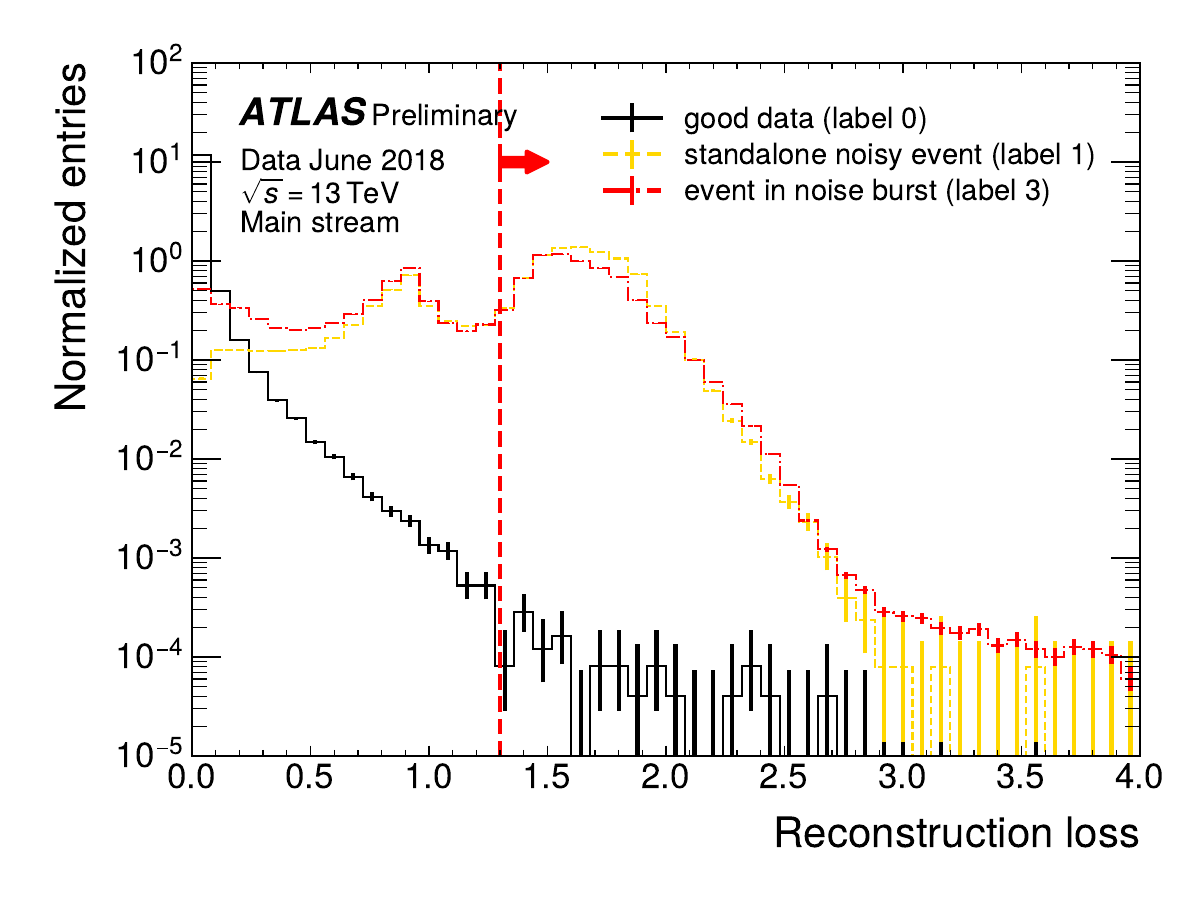}
        \caption{AE reconstruction loss distributions for different events as categorized by reference noise burst algorithm. The chosen anomaly threshold \(L_\mathrm{th}=1.3\) separates good data from the bulk of the noisy events. Figure taken from Ref.~\cite{ATL-DAPR-PUB-2024-002}.}
        \label{fig:loss-distribution}
    \end{subfigure}
    \caption{Illustrations of LAr noise bursts and their reconstruction loss distributions.}
    \label{fig:noise-loss-combined}
\end{figure}

The LAr calorimeter periodically suffers from bursts of coherent noise, which were found to intensify with increasing luminosity. Such bursts typically last for no longer than a millisecond, which is still a considerable period given the LHC bunch crossing rate of 25 ns. Figure~\ref{fig:noise-burst} shows an example of a typical event with a total energy of 2 TeV recorded during a noise burst. A well-established algorithm to identify such noisy events already exists within \mbox{ATLAS} and was therefore used to provide ground-truth labels to benchmark the AE performance. Figure~\ref{fig:loss-distribution} shows that the AE reconstruction loss for the events identified as noisy by this algorithm is significantly higher compared to the rest. Based on this distribution, the threshold \(L_\mathrm{th} = 1.3\) was chosen to classify events as anomalous. An anomalous time window is obtained by requiring several such consecutive events in order to reduce the risk of single rare events being flagged as detector issues, as is done for the reference algorithm.

\begin{figure}[htb!]
    \centering
    \includegraphics[width=0.7\linewidth]{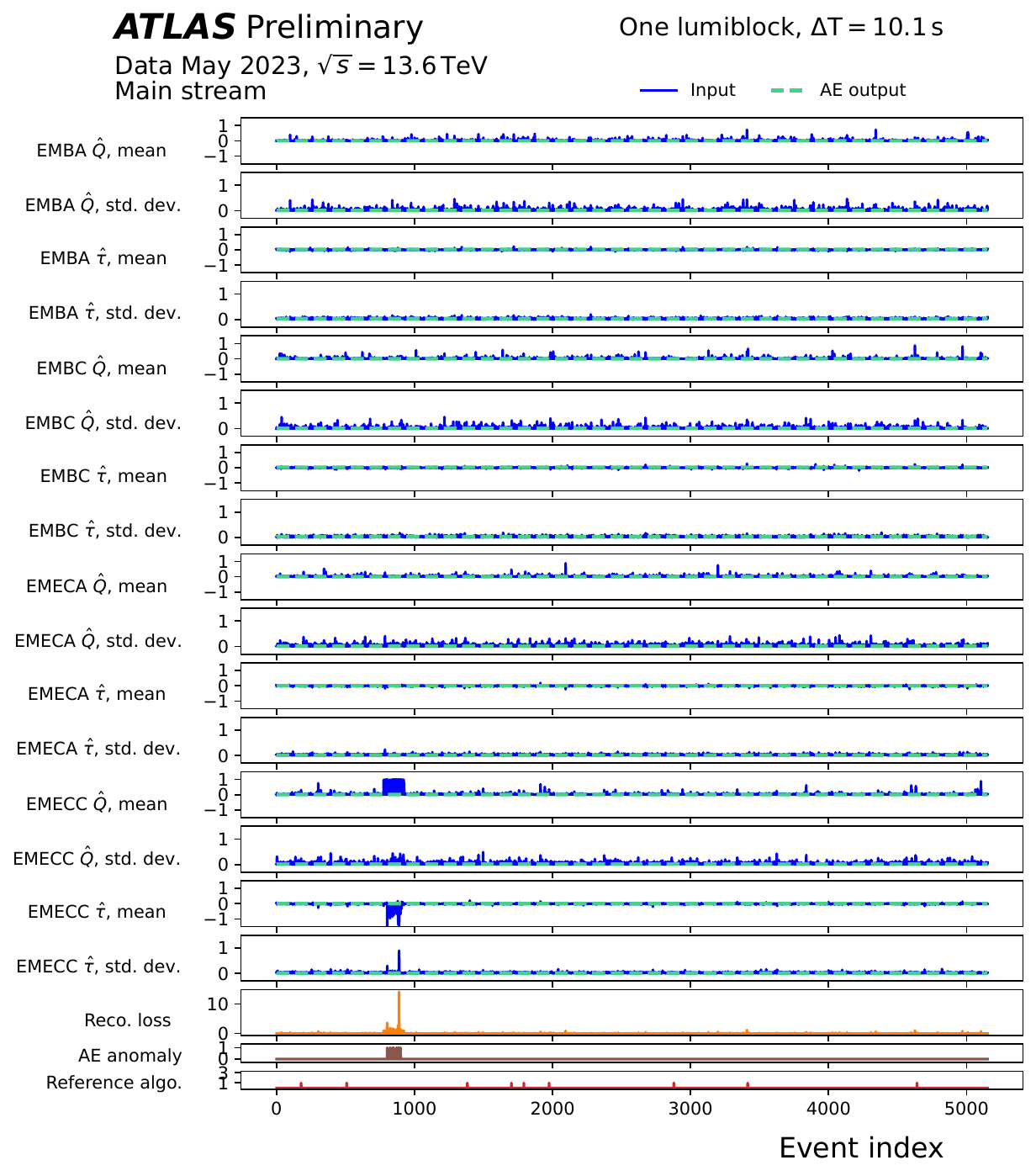}
    \caption{Original (blue) and AE-reconstructed (cyan) time series for calorimeter partitions (top 16 rows). The bottom three rows show the reconstruction loss, \(L\), events with \(L > 1.3\), and noisy event labels from a reference algorithm. An anomalous window near event index 900 is flagged by the AE but not by the reference algorithm. Figure taken from Ref.~\cite{ATL-DAPR-PUB-2024-002}.}
    \label{fig:anomaly-3}
\end{figure}

Although the AE was primarily validated using noise burst-enriched data, its application remains more general. The AE complements the reference algorithm, occasionally identifying additional anomalous time windows. One such example is shown in Figure~\ref{fig:anomaly-3} for proton--proton collision data, likely corresponding to a temporary failure of a vacuum diffusion pump. The plot shows that the AE reconstruction loss passes the anomaly threshold \(L > L_\mathrm{th}\) for a cluster of events and thus flags the issue, unlike the reference algorithm. Additional studies in Ref.~\cite{ATL-DAPR-PUB-2024-002} show that the model also performs well on cosmic or heavy--ion collision data.

\section{Conclusion}
\label{sec:conclusion}

A proof-of-principle time series anomaly detection method based on an LSTM autoencoder has been developed for the ATLAS Liquid Argon calorimeter. The algorithm is capable of identifying brief anomalous periods in the calorimeter time series, complementing existing data quality monitoring techniques. Validation studies show that the autoencoder successfully reconstructs good-quality data while identifying known anomalous events. Moreover, the method generalizes to previously unseen detector issues not flagged by existing algorithms. These results illustrate the potential of unsupervised anomaly detection as an early warning system for rare detector problems, paving the way for more robust and automated LAr calorimeter monitoring.

\section*{Acknowledgements}
% Acknowledgements should follow immediately after the conclusion.

% TODO: include funding information
\paragraph{Funding information}
This contribution is part of a project that has received funding from the European Research Council (ERC) under the European Union’s Horizon 2020 research and innovation programme (Grant agreement No. 948254).
% Authors are required to provide funding information, including relevant agencies and grant numbers with linked author's initials. Correctly-provided data will be linked to funders listed in the \href{https://www.crossref.org/services/funder-registry/}{\sf Fundref registry}.

\bibliography{bibliography.bib}

\end{document}